\documentclass[%
 reprint,
superscriptaddress,
preprintnumbers,
nofootinbib,
 amsmath,amssymb,
 aps,
 prl,
longbibliography
]{revtex4-1} 

\usepackage[pdftex]{graphicx}
\usepackage[utf8]{inputenc}
\usepackage{dcolumn}
\usepackage{bm}
\usepackage[colorlinks,allcolors=blue,linktocpage]{hyperref}
\usepackage{cleveref}

\begin{document}
\title{Uniting low-scale leptogeneses}
\author{Juraj Klarić}
\author{Mikhail Shaposhnikov}
\author{Inar Timiryasov}
\affiliation{Institute of Physics, Laboratory for Particle Physics and Cosmology,
École polytechnique fédérale de Lausanne,
CH-1015 Lausanne,
Switzerland}


\begin{abstract}

In this work we demonstrate that what was previously considered as different mechanisms of baryon asymmetry generation involving
two right-handed Majorana neutrinos with masses far below the GUT scale--- leptogenesis via neutrino oscillations and resonant leptogenesis---are actually united.
We show that the observed baryon asymmetry can be generated for all experimentally allowed values of the right-handed neutrino masses above $M_N \gtrsim 100$ MeV.
Leptogenesis is effective in a broad range of the parameters, including mass splitting between two right-handed neutrinos as big as $\Delta M_N/M_N \sim 0.1$,
as well as mixing angles between the heavy and light neutrinos large enough to be accessible to planned intensity experiments or future colliders.
\end{abstract}

\maketitle

\paragraph{Introduction.} 
\label{par:introduction}
Flavor oscillations of neutrinos is the only laboratory tested phenomenon pointing on
the incompleteness of the Standard Model (SM).
The presence of the ordinary baryonic matter in the observed amounts cannot be explained within the SM as well (see, e.g. review~\cite{Canetti:2012zc}).
The minimal renormalisable extension of the SM contains two or more gauge singlet right-handed neutrinos which allow for a Dirac mass matrix $m_D$ for the neutrinos.
These singlet right-handed neutrinos are the only particles which can have Majorana masses with the mass matrix $M_M$.
Quantum field theory suggests that these mass terms---like any other coefficients in front of renormalisable operators---should be determined experimentally.
Remarkably, diagonalising the common neutrino mass matrix  one finds that if $M_M \gg m_D$, the mass matrix of left-handed neutrinos is $m_\nu \simeq - m_D^2/M_M$. This is the famous seesaw formula~\cite{Minkowski:1977sc,Yanagida:1979as,GellMann:1980vs,Mohapatra:1979ia,Schechter:1980gr,Schechter:1981cv}.
An important consequence of the theory is the mixing between the light neutrinos and the heavier ones.
This mixing allows the right-handed neutrinos to interact with the rest of the SM, so from the experimental point of view they behave like \emph{heavy neutral leptons} (HNLs).
The search for HNLs is an important part of physics programs of most accelerator experiments,
both operating~\cite{Liventsev:2013zz,Aaij:2014aba,Artamonov:2014urb,Aad:2015xaa,Khachatryan:2015gha,Sirunyan:2018mtv,Boiarska:2019jcw} and planned~\cite{Alekhin:2015byh,Gligorov:2017nwh,Feng:2017uoz,Kling:2018wct,Drewes:2018gkc,Strategy:2019vxc,Abada:2019lih}.
The capability of explaining neutrino masses strongly motivates HNL searches.
However, there are other intriguing consequences of the theory outlined above.
Yukawa couplings of right-handed neutrinos can carry new sources of $CP$ violation, while HNLs themselves deviate from equilibrium in one way or another.
Sphaleron processes in the early Universe provide violation of the baryon number~\cite{Kuzmin:1985mm}. Therefore the Sakharov conditions can be satisfied and generation of the Baryon Asymmetry of the Universe (BAU) is possible.
HNLs interact only with leptons, so it is the lepton asymmetry which is generated and transferred to the baryon sector by the sphaleron processes. This mechanism is known as \emph{leptogenesis}.\footnote{Let us note in passing that transfer of asymmetry from the lepton sector is efficient at temperatures exceeding $\simeq 130$~GeV~\cite{DOnofrio:2014rug}.
This means that  HNLs responsible for leptogenesis serve as a unique probe of the very early Universe.}
The suggestion along these lines was proposed by Fukugita and Yanagida~\cite{Fukugita:1986hr} who considered very heavy right-handed neutrinos with masses above $10^9$~GeV~\cite{Davidson:2002qv}.
The mass scale of leptogenesis can be significantly lowered if two HNLs are nearly degenerate in mass, this phenomenon was dubbed \emph{resonant leptogenesis}~\cite{Liu:1993tg,Flanz:1994yx,Flanz:1996fb,Covi:1996wh,Covi:1996fm,Pilaftsis:1997jf,Buchmuller:1997yu,Pilaftsis:2003gt}.
Later it was realized that GeV-scale right-handed neutrinos can also generate the BAU in \emph{leptogenesis via oscillations}~\cite{Akhmedov:1998qx,Asaka:2005pn}
(for more recent work see e.g.~\cite{Shaposhnikov:2006nn,Shaposhnikov:2008pf,Canetti:2010aw,Asaka:2010kk,Anisimov:2010gy,Asaka:2011wq,Besak:2012qm,Canetti:2012vf,Drewes:2012ma,Canetti:2012kh,Shuve:2014zua,Bodeker:2014hqa,Abada:2015rta,Hernandez:2015wna,Ghiglieri:2016xye,Hambye:2016sby,Hambye:2017elz,Drewes:2016lqo,Asaka:2016zib,Drewes:2016gmt,Hernandez:2016kel,Drewes:2016jae,Asaka:2017rdj,Eijima:2017anv,Ghiglieri:2017gjz,Eijima:2017cxr,Antusch:2017pkq,Ghiglieri:2017csp,Eijima:2018qke,Ghiglieri:2018wbs,Ghiglieri:2019kbw,Ghiglieri:2020ulj}).
Both scenarios require two HNLs with nearly degenerate masses.\footnote{
The mass degeneracy of two HNLs is an interesting feature from the theoretical point of view
as it may be a result of a global leptonic symmetry - in this case a pair of Majorana neutrinos $N$ can be joined into a quasi-Dirac fermion.
An interesting feature is that it also allows for sizable mixings $\Theta_{\alpha I}$ in a technically natural way~\cite{Wyler:1982dd,Mohapatra:1986bd,Branco:1988ex,GonzalezGarcia:1988rw,Shaposhnikov:2006nn,Kersten:2007vk,Abada:2007ux,Gavela:2009cd}.}
The absence of a preferred mass scale of leptogenesis calls for a vast and diverse search program. High intensity frontier experiments, especially SHiP~\cite{Alekhin:2015byh}, provide an unparalleled opportunity if $M$ is in a few GeV region, whereas future colliders, such as FCC-ee~\cite{Blondel:2014bra,Abada:2019lih,Antusch:2016ejd,Antusch:2016vyf}, or CEPC~\cite{Antusch:2016ejd,Antusch:2016vyf} will cover a significant portion of the parameter space of heavier HNLs.

\paragraph{Resonant leptogenesis and leptogenesis via oscillations.} 
\label{par:RLGvsOSC}
After inflation the baryon and lepton numbers of the Universe as well as the number of HNLs may well be zero, and we will assume that this is indeed the case  \cite{Bezrukov:2008ut}\footnote{This is not necessarily so if the $\nu$MSM is supplemented by higher dimensional operators \cite{Bezrukov:2011sz,Shaposhnikov:2020aen}.}.
The baryon asymmetry of the Universe in both \emph{leptogeneses} is produced in a set of  processes including  scatterings, decays, coherent oscillations of HNLs, and anomalous sphaleron transitions.

The conceptual difference between the two leptogeneses is the moment in the history of the Universe when the asymmetry is generated.
In resonant leptogenesis the BAU is generated when the temperature drops below the heavy neutrino mass, $T \lesssim M_N$, and the neutrinos begin to decay out of equilibrium.
As conversion between lepton and baryon number requires fast electro-weak sphaleron processes -- this implies a lower bound on the heavy neutrino masses around ${M_N \sim T_\text{sph} \simeq 130}$~GeV~\cite{DOnofrio:2014rug}.
Indeed, this is close to the lowest heavy neutrino mass for which resonant leptogenesis was studied in~\cite{Pilaftsis:2005rv}.
On the other hand, in baryogenesis via neutrino oscillations, the BAU is primarily produced during the equilibration of the heavy neutrinos.
It has been argued that baryogenesis via oscillation only works when $M_N$ is below $M_W$~\cite{Blondel:2014bra},
since the equilibration rate of the heavy neutrinos generically exceeds the Hubble rate when $M_N \sim T$, as the neutrinos are become heavy enough to decay into $W$ and $Z$ bosons.
One simply arrives at the conclusion that these are two genuinely different mechanisms of leptogenesis.
In this \emph{letter} we show for the first time that this is not the case, and that leptogenesis with two HNLs is operative for all values $M_N$ larger than a fraction of GeV.

To avoid confusion with terminology of oscillations and resonances (present in both mechanisms), in the remainder of the text, we borrow the language often used for dark matter production mechanisms, and refer to the two mechanisms as:
\emph{freeze-in} leptogenesis, which corresponds to leptogenesis via oscillations, where the BAU is mainly generated during the production of the HNLs;
and \emph{freeze-out} leptogenesis, which corresponds to conventional resonant leptogenesis, where the majority of the BAU is generated during their out-of equilibrium decays.

\paragraph{A unified picture.} 
\label{par:unified_picture}

The first question one may ask when comparing the two mechanisms is whether the equations governing
the production of the BAU are the same.
There have been several approaches to deriving the evolution equations for resonant leptogenesis and leptogenesis via oscillations.
In the case of resonant leptogenesis the perturbative computation leads to a divergent heavy neutrino decay asymmetry in the limit of exactly degenerate heavy neutrinos, see, e.g.~\cite{Liu:1993tg}.
This can be understood as a breakdown of the usual perturbation theory,
since the unstable heavy neutrinos cannot appear as asymptotic $S$-matrix states.
After the initial developments~\cite{Pilaftsis:1997jf, Pilaftsis:1998pd,Liu:1993tg,Flanz:1996fb,Covi:1996fm,Pilaftsis:1997dr,Buchmuller:1997yu},
the studies of resonant leptogenesis have taken a more formal turn with the goal of deriving the evolution equations from first principles, in particular using methods from non-equilibrium QFT,
in particular the \emph{closed-time-path} (CTP) formalism~\cite{Buchmuller:2000nd,DeSimone:2007edo,DeSimone:2007gkc,Garny:2009rv,Garny:2009qn,Anisimov:2010aq,Beneke:2010wd,Garny:2011hg,Iso:2013lba,Garbrecht:2011aw,Herranen:2010mh,Fidler:2011yq, Herranen:2011zg, Millington:2012pf,Millington:2013isa,Dev:2014laa,Dev:2014tpa,Dev:2015wpa}.
For leptogenesis via neutrino oscillations, where the neutrinos are close to relativistic, the equations are often derived by generalizing the treatment of Sigl and Raffelt~\cite{Sigl:1992fn} of relativistic mixed neutrinos to the scenario with additional heavy states~\cite{Akhmedov:1998qx, Asaka:2005pn}.
The same type of equations can be derived in the CTP formalism~\cite{Garbrecht:2011aw}
 if we assume a common mass shell for the two heavy neutrinos.
This approach has successfully been used in studies of both resonant leptogenesis~\cite{Garbrecht:2014aga} and leptogenesis via neutrino oscillations~\cite{Drewes:2016gmt}, by taking the non-relativistic and relativistic limits respectively.
The importance of non-relativistic corrections to leptogenesis via oscillations was pointed out in~\cite{Hambye:2016sby,Eijima:2017anv,Ghiglieri:2017gjz}.
The equations that we use in the remainder of this work are a generalization of the ones used in~\cite{Eijima:2017anv,Eijima:2018qke} to the non-relativistic case (cf~refs.~\cite{Ghiglieri:2019kbw, Bodeker:2019rvr}),
and are consistent with the equations derived for resonant leptogenesis~\cite{Garbrecht:2011aw}:
\begin{subequations}
  \begin{align}
    i \frac{d n_{\Delta_\alpha}}{dt}
    &= -2 i \frac{\mu_{\alpha}}{T} \int \frac{d^{3} k}{(2 \pi)^{3}} \operatorname{Tr}\left[\Gamma_{\alpha}\right] f_{N}\left(1-f_{N}\right) \nonumber\\
    &\quad +i \int \frac{d^{3} k}{(2 \pi)^{3}} \operatorname{Tr}\left[\tilde{\Gamma}_{\alpha}\left(\bar{\rho}_{N}-\rho_{N}\right)\right],
    \label{kin_eq_a}
    \\
    i \, \frac{d\rho_{N}}{dt}
    &= \left[H_{N}, \rho_{N}\right]-\frac{i}{2}\left\{\Gamma, \rho_{N}-\rho_{N}^{eq} \right\} \nonumber\\
    &\quad-\frac{i}{2} \sum_{\alpha} \tilde{\Gamma}_{\alpha}\left[2 \frac{\mu_{\alpha}}{T} f_{N}\left(1-f_{N}\right)\right] ,
    \label{kin_eq_b}
    \\
    i \, \frac{d \bar{\rho}_{N}}{d t}
    &= -\left[H_{N}, \bar{\rho}_{N}\right]-\frac{i}{2}\left\{\Gamma, \bar{\rho}_{N}-\rho_{N}^{eq} \right\} \nonumber\\
    &\quad+\frac{i}{2} \sum_{\alpha} \tilde{\Gamma}_{\alpha}\left[2 \frac{\mu_{\alpha}}{T} f_{N}\left(1-f_{N}\right)\right],
    \label{kin_eq_c}
  \end{align}\label{kin_eq}\end{subequations}
  where $n_{\Delta_\alpha} \equiv L_\alpha - B/3$ are the lepton asymmetries which can be related to the chemical potentials through the susceptibility matrix $\mu_\beta = \omega_{\alpha \beta} n_{\Delta_\alpha}$, and $\rho_N$ and $\bar{\rho}_{N}$ are the matrices of the heavy neutrino number densities.
The equations are governed by the equilibration matrices ${\Gamma=\sum_\alpha \Gamma_\alpha}$ and ${\tilde{\Gamma}=\sum_\alpha \tilde{\Gamma}_\alpha}$, the effective Hamiltonian $H_N$ describing the neutrino oscillations and  ${\rho_{N}^{eq} = \mathbf{1}_{2\times 2}\cdot f_N}$, where $f_N$ is the equilibrium distribution of the HNLs.

Equations~\eqref{kin_eq} describe both leptogeneses. At the same time, equations derived in ref.~\cite{Dev:2014laa,Garbrecht:2014aga} for the case of resonant leptogenesis have a similar form except for the fact that the equations for $\bar{\rho}_{N}$ are are not independent from those for $\rho_{N}$ which is not the case in \cref{kin_eq}. 
However, there is no contradiction since in the non-relativistic limit \cref{kin_eq_c}  indeed becomes a conjugate of \cref{kin_eq_b}.\footnote{Another important distinction is that the equations from ref.~\cite{Dev:2014laa} contain the so-called \emph{effective} Yukawa couplings~\cite{Pilaftsis:1997jf,Pilaftsis:2003gt}. Their purpose is to remedy the breakdown of the density matrix description when the heavy neutrino energy differences become hierarchical.
However, since we focus our study on the quasi-degenerate regime of leptogenesis, we assume these effects may be neglected~\cite{Kartavtsev:2015vto,Dev:2017wwc}.
}
The rates entering \cref{kin_eq} pose the main theoretical challenge.
A lot of effort has been made to compute them at high temperatures~\cite{Anisimov:2010gy,Besak:2012qm,Ghisoiu:2014ena,Garbrecht:2013bia,Biondini:2017rpb,Garbrecht:2019zaa},
however, the rates in the literature are typically helicity-averaged.
For relativistic HNLs the rate is helicity-dependent and requires a more careful calculation~\cite{Ghiglieri:2017gjz,Ghiglieri:2017csp,Ghiglieri:2016xye}.
The helicity-dependent rates have only been calculated in the relativistic limit, and cannot be applied in the intermediate regime, which is crucial to connect the two mechanisms.
In ref.~\cite{long_paper} we approximate the rate $\Gamma$ and show that the results are insensitive to the details of such estimates.

\paragraph{Parameter space of leptogenesis.} 
\label{par:parameter_space_of_leptogenesis}
The system of equations~\eqref{kin_eq} needs to be solved numerically to obtain an accurate estimate of the BAU.
Solving momentum-averaged equations (see~\cite{Asaka:2011wq, Ghiglieri:2017csp}), we perform a parameter scan over the masses and mixing angles consistent with the observed light neutrino masses using the Casas-Ibarra parametrization~\cite{Casas:2001sr}.

\begin{figure}[h]
    \centering
    \includegraphics[width=0.4\textwidth]{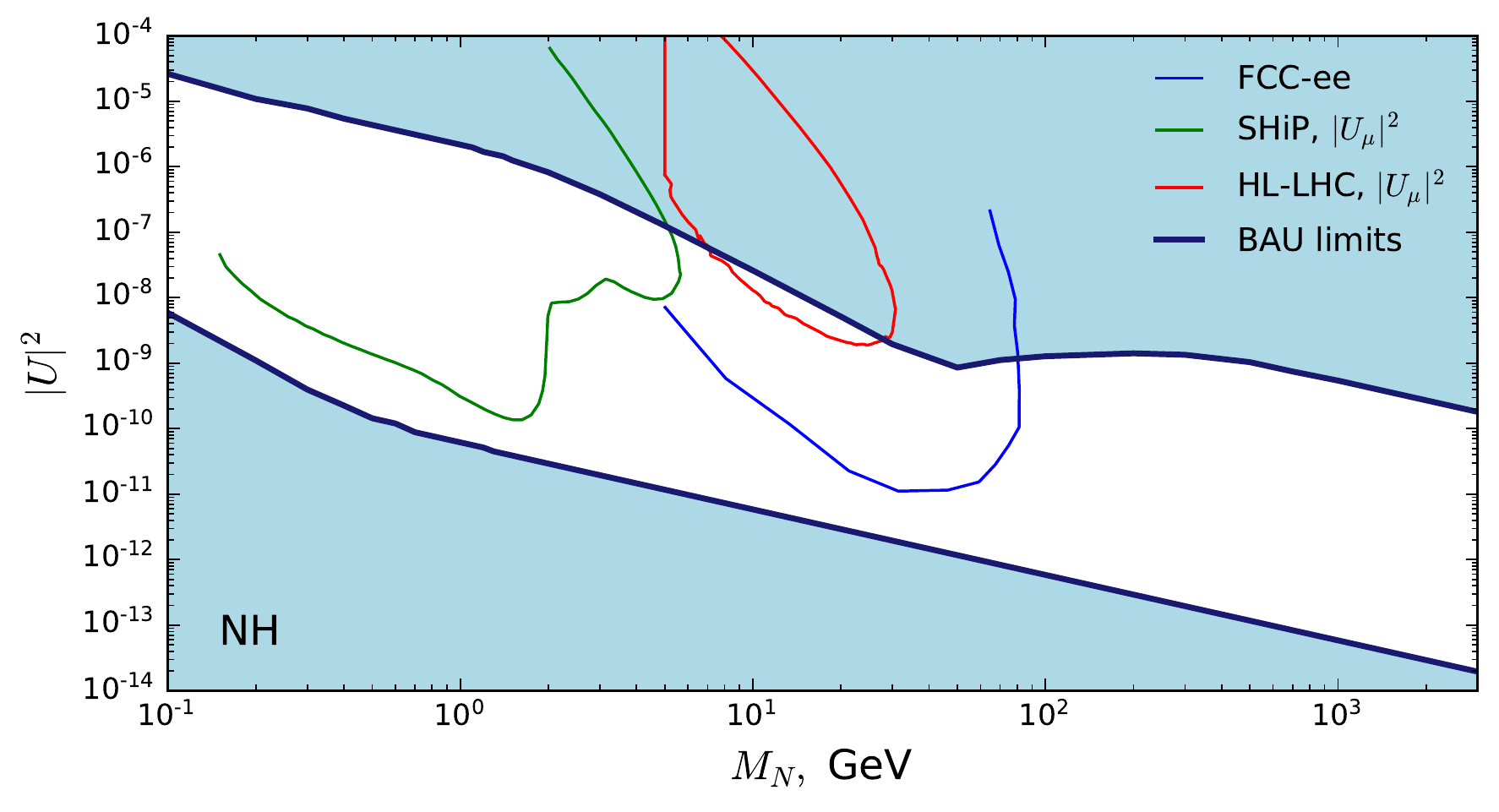}
    \includegraphics[width=0.4\textwidth]{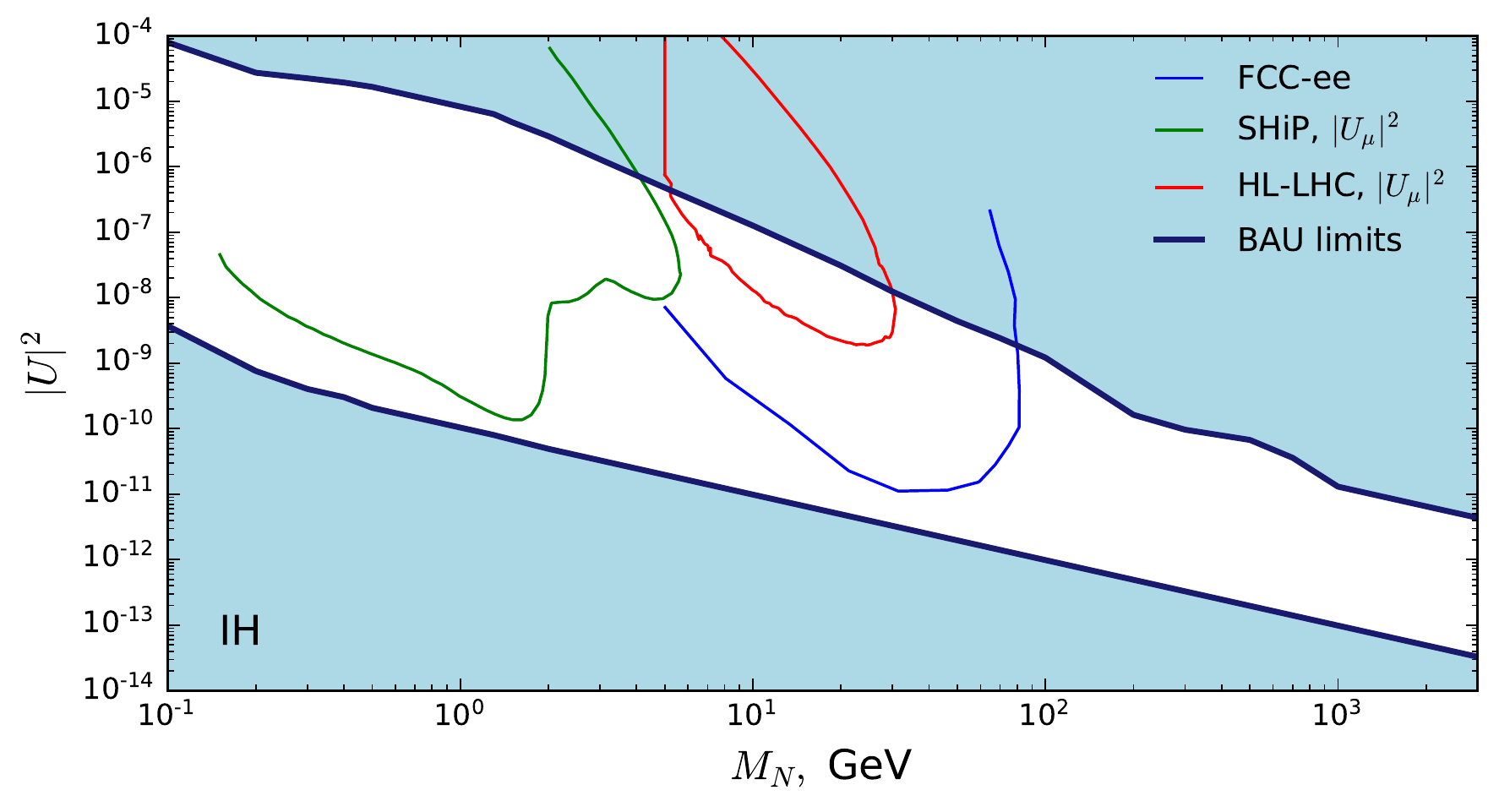}
    \caption{Within the white regions it is possible to reproduced the observed
value of the BAU. 
Upper panel: Normal hierarchy. Lower panel: inverted hierarchy. 
For comparison we also show the expected sensitivities of SHiP (green), HL-LHC (red) and FCC-ee (blue),
as representative experiments in their corresponding mass range.
The sensitivitiy lines are taken from~\cite{Alekhin:2015byh,Boiarska:2019jcw,Antusch:2016ejd}.
}
    \label{fig:results}
\end{figure}

The neutrino flavor eigenstates can be expressed as
    \;$\nu_\alpha = U_{\alpha i} \nu_i + \Theta_{\alpha I} N_I^c,$\;
where $\nu_i$ and $N_I$ are light and heavy mass eigenstates with masses $m_i$ and $M_I$ respectively,
$U_{\alpha i}$ is the PMNS matrix and $\Theta_{\alpha I}$ is the mixing between active neutrinos and HNLs.
Here we consider the case of two HNLs\footnote{The third HNL---if it exists---could be light and very weakly coupled~\cite{Shaposhnikov:2006nn}, which makes it a perfect dark matter candidate as it the case in the $\nu$MSM~\cite{Asaka:2005an,Asaka:2005pn,Canetti:2012kh,Canetti:2012vf,Ghiglieri:2019kbw}.} which is compatible with the neutrino oscillation data, so $I = 1,2$ and ${M_{1,2} = M \pm \Delta M}$.
It is convenient to characterize the overall strength of the mixing using
$|U|^2 = \sum_{\alpha I} |\Theta_{\alpha I}|^2$. The see-saw requires that $|U|^2\geq \sum_\alpha m_\alpha / M$, whereas demanding successful leptogenesis sets up an upper bound on $|U|^2$. In \cref{fig:results} we show the region in the parameter space where the observed value of the BAU can be generated. As one can see, the results depend on the neutrino mass hierarchy.\footnote{In the case of two HNLs which we consider here, the lightest active neutrino is almost massless and the neutrino mass spectrum is hierarchical.}
One can show~\cite{long_paper} that
the allowed region extends to heavier masses and both upper and lower bounds scale as
$ |U|^2 \propto 1/ M$. This scaling breaks down around $M\sim10^7$~GeV due to flavor effects~\cite{Abada:2006fw, Nardi:2006fx, Abada:2006ea, Blanchet:2006be, Pascoli:2006ie, DeSimone:2006nrs}, as well as the maximal mass splitting becoming of order $\Delta M/ M \sim \mathcal{O}(1)$,
which leads to a breakdown of the quasiparticle approximation used to derive the quantum kinetic equations.
As one can see in \cref{fig:results}, there is a continuous region in the $U^2-M$ plane
where leptogenesis in its seemingly different incarnations is operative.

\begin{figure}[h]
    \centering
    \includegraphics[width=0.4\textwidth]{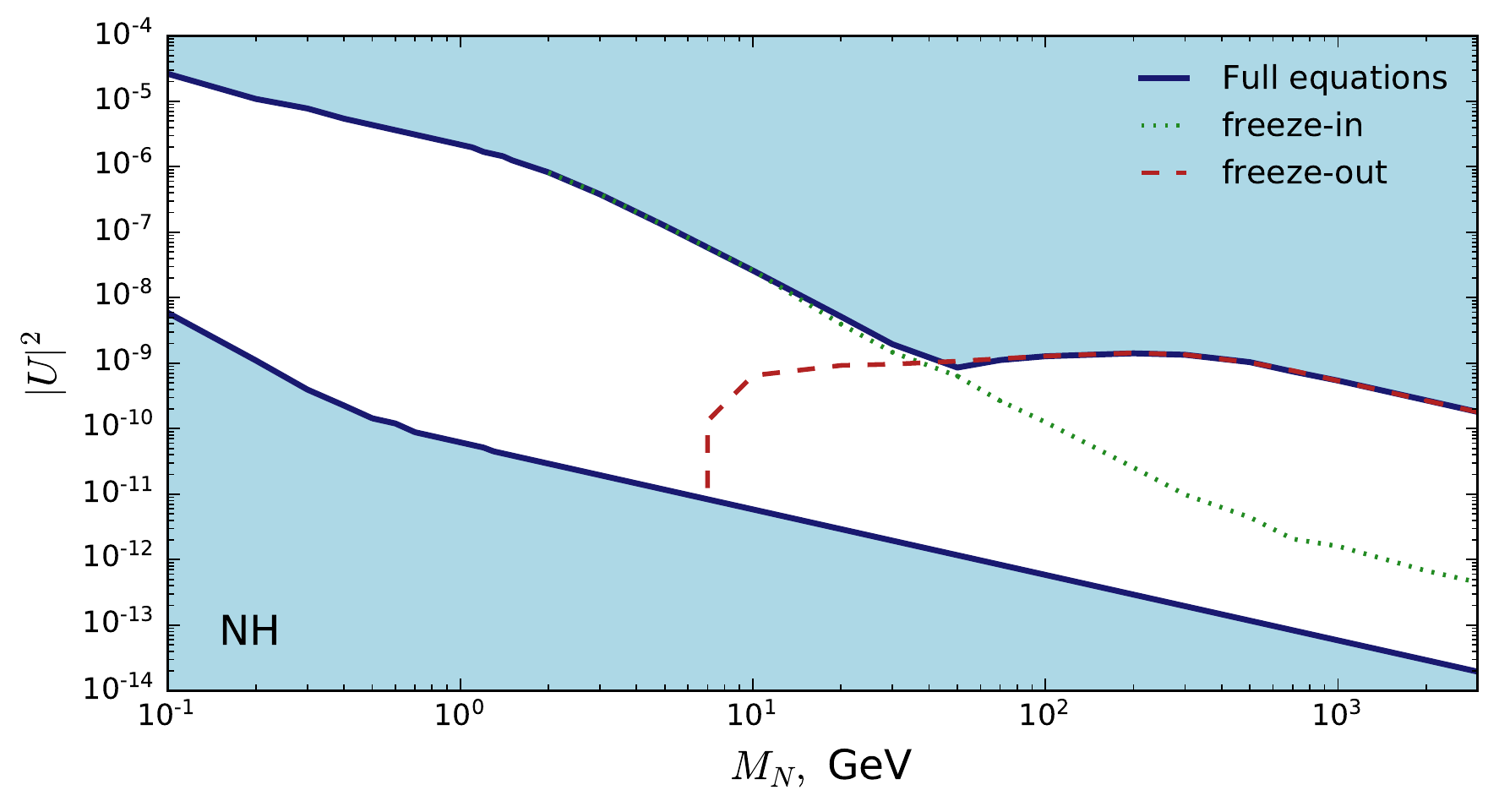}
    \caption{
    Regions of parameter space corresponding to the freeze-in regime (no source term), green dotted line, and to the freeze-out regime (thermal initial conditions), red dashed line. Together the two regimes span the whole low-scale leptogenesis parameter space.
    It is interesting to note that freeze-in leptogenesis remains viable up to arbitrarily large masses, albeit for mixing angles close to the seesaw scale. The NH case is shown; the similar pattern is observed for the IH.}
    \label{fig:different_regimes}
\end{figure}

\paragraph{Regimes of leptogenesis.} 
\label{par:regimes_of_leptogenesis}
As we can see from \cref{fig:results}, there is no clear separation between the two leptogeneses.
We distinguish between them based on when the majority of the asymmetry is generated, i.e. during freeze-in or freeze-out.
To fully separate these regimes, we consider different initial conditions for the heavy neutrinos.
For the freeze-out parameter space we start with thermalized heavy neutrinos, and rely purely on their out-of equilibrium decays.
Similarly, for freeze-in leptogenesis, we artificially turn off the terms driving the heavy neutrinos out of equilibrium.
Of course, the physical solution relies on the presence of both effects.
The comparison between these three ``parameter spaces'' is shown in figure~\cref{fig:different_regimes}.

Perhaps surprisingly, we find that both regimes extend beyond the masses we would naively associate with them.
Freeze-in leptogenesis extends far beyond $M_W$, and freeze-out leptogenesis is possible already for masses as low as $5$ GeV.\footnote{GeV-scale freeze-out leptogenesis was already studied in~\cite{Hambye:2016sby},
however, using the usual Boltzmann equations which are not appropriate in this mass regime.}
This statement can be quantified in the following way.
If one starts from the thermal initial conditions for HNLs, then only freeze-out can contribute. This is shown by the red dashed line in \cref{fig:different_regimes}. On the other hand, we can set to zero the time derivative of the equilibrium distribution $\rho_{N}^{e q}$, which we refer to as a source term. In this case there is no deviation from equilibrium during freeze-out and all asymmetry is generated during freeze-in, see the green dotted line in \cref{fig:different_regimes}.
The main ingredients which make the overlap of these regimes possible are: \emph{(i)} flavor hierarchical washout; \emph{(ii)} deviation from the equilibrium due to the expansion of the Universe; \emph{(iii)} approximate lepton number conservation.

\begin{figure}[h!]
    \centering
    \includegraphics[width=0.4\textwidth]{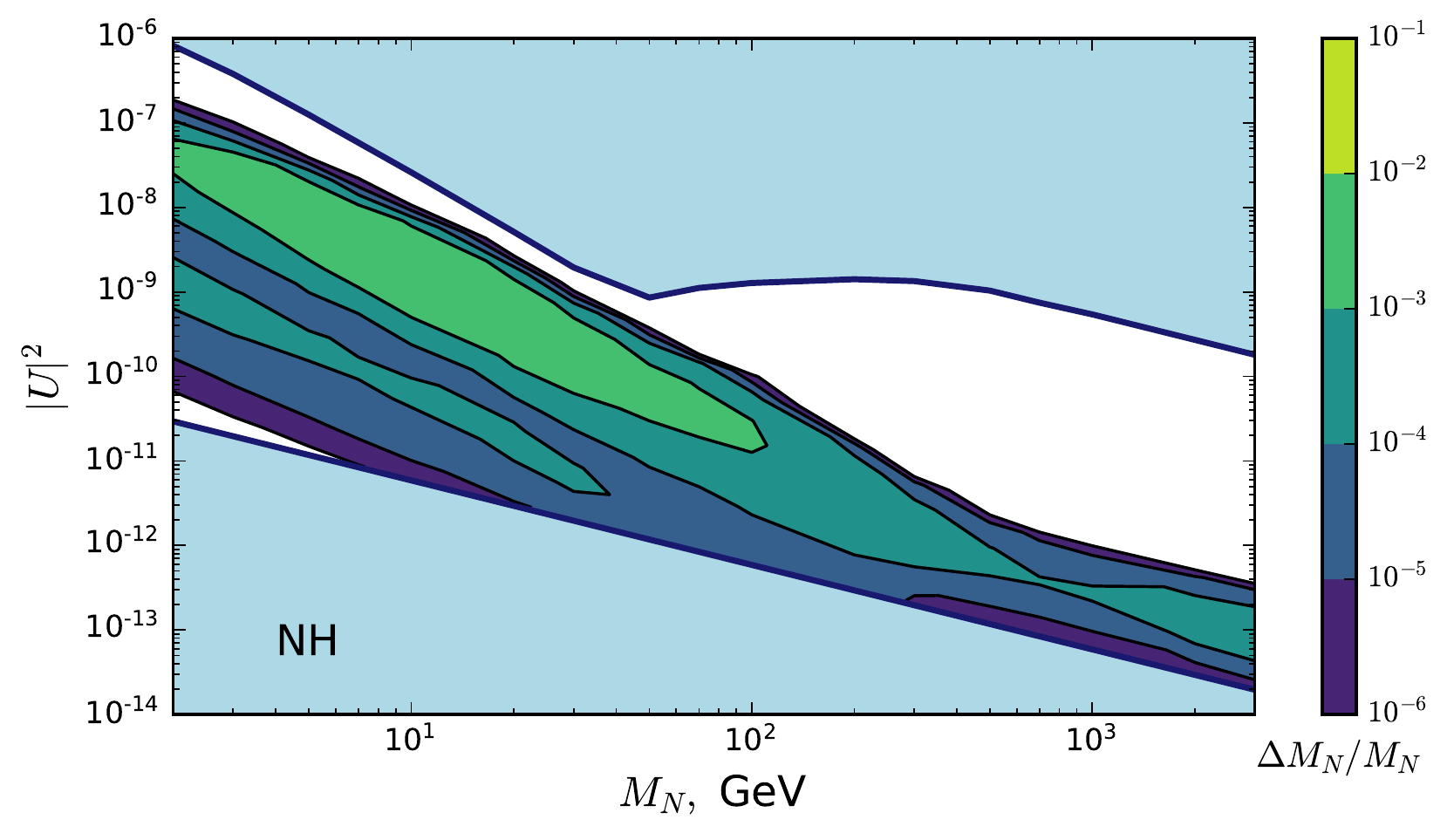}
    \includegraphics[width=0.4\textwidth]{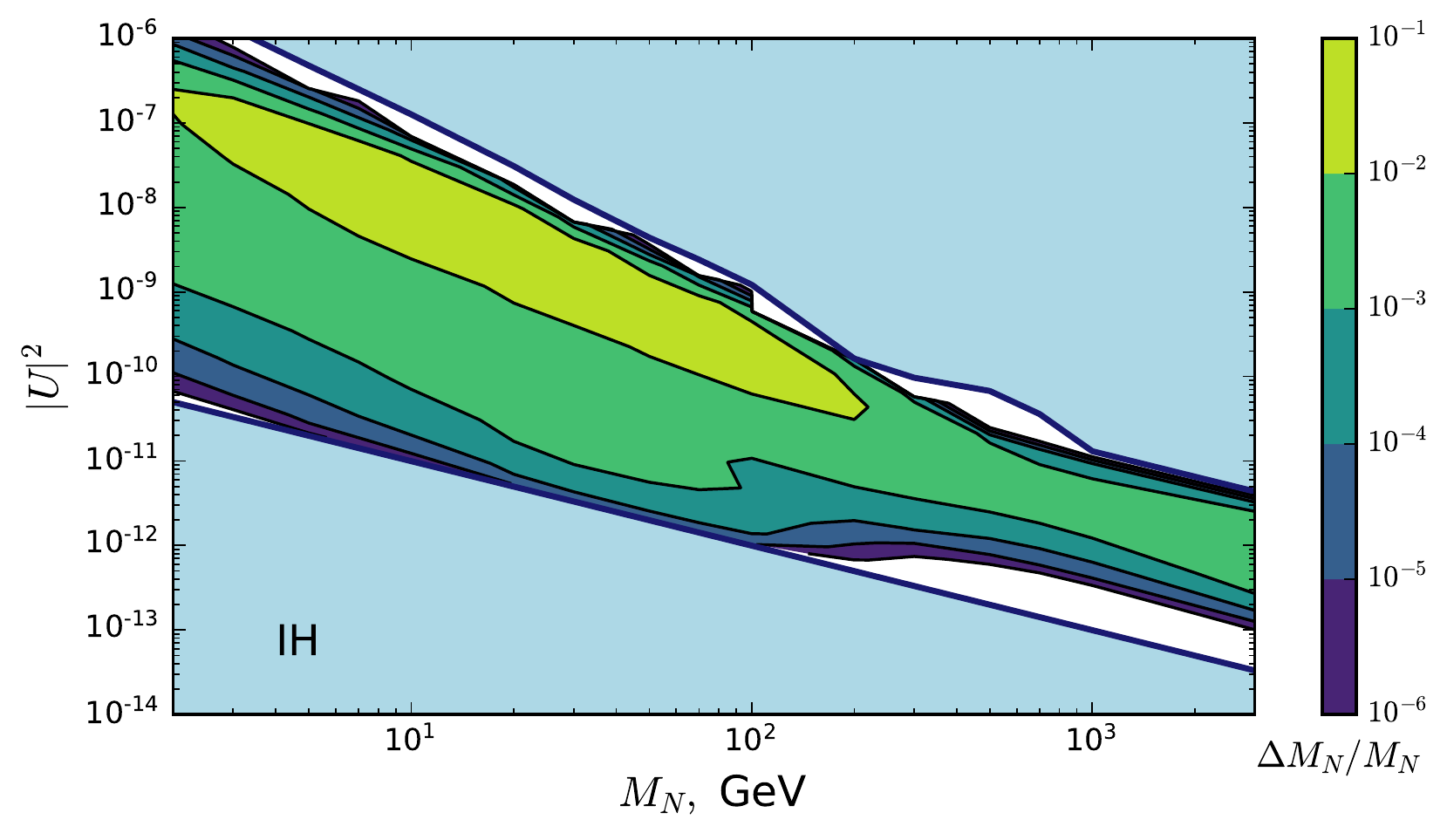}
    \caption{The maximal mass splitting consistent with leptogenesis for fixed $M$ and $U^2$. The white region corresponds to mass splittings below $10^{-6}$.
    It is interesting to see that the region of large mass splittings mostly coincides with the freeze-in leptogenesis regime.
This can be expected, as for large mass splitting the majority of the BAU is generated at high temperatures, before the HNLs begin to decay.}
    \label{fig:massSplitting}
\end{figure}

When the heavy neutrino masses are of the same order as the temperature, the ratio of the equilibration and Hubble rates is in general quite large,
with the smallest value for normal hierarchy around $\mathcal{O}(30)$.
Naively this would lead us to expect that any asymmetries generated during freeze-in would be erased by the strong washout.
However, the washout rate of a particular lepton flavor can be several orders of magnitude smaller than the equilibration rate for the heavy neutrinos.
The presence of a \emph{flavor hierarchical washout} is almost completely determined by the $CP$-violating phase $\delta$ and the Majorana phases from the Pontecorvo-Maki-Nakagawa-Sakata (PMNS) matrix as parametrised in~\cite{Zyla:2020zbs}.
It can range from $\mathcal{O}(10^{-3})$ to $\mathcal{O}(10^{-1})$ for NH, while it can be as small as $\mathcal{O}(10^{-4})$, or completely non-hierarchical in the case of IH.
For large masses of HNLs, freeze-in leptogenesis crucially depends on the presence of such hierarchies (cf.~\cite{Garbrecht:2014bfa}, where the importance of a hierarchical washout was pointed out in the 3 HNL case).
Furthermore, we find that freeze-in is the dominant mechanism when the mass splitting between the heavy neutrinos is sizable $\Delta M_N / M_N \sim \mathcal{O}(10^{-2})$, as demonstrated in fig.~\ref{fig:massSplitting}.

At the same time, we find successful freeze-out leptogenesis at the few GeV-scale. The main reason behind this effect is that the decay asymmetries of the heavy neutrinos can be close to $\mathcal{O}(1)$.
The deviation from equilibrium caused by the heavy neutrino freeze-out in such a scenario will be suppressed by $10^{-3}M^2/T^2$, and can still lead to the observed baryon asymmetry.

Finally, we also find that even in the absence of flavor hierarchical washout, large mixing angles remain viable for heavy neutrino masses above $M_W$.
The main reason behind this observation is the presence of an \emph{approximately conserved lepton number}.
If the pair of heavy Majorana neutrinos is close to degenerate in mass, they can be combined into a single pseudo-Dirac neutrino which can carry a lepton number.
This type of scenario was studied as a technically natural way of adding light right-handed neutrinos to the SM~\cite{Wyler:1982dd,Mohapatra:1986bd,Branco:1988ex,GonzalezGarcia:1988rw,Shaposhnikov:2006nn,Kersten:2007vk,Abada:2007ux,Gavela:2009cd}.
However, the importance of an approximate lepton number in preventing large washout during leptogenesis was first noted in~\cite{Blanchet:2009kk}.
The small parameter determining the conservation of this lepton number is the ratio of the heavy neutrino mass splitting and their interaction (decay) rate.

\paragraph{Discussion and conclusions.} 
\label{par:discussion_and_conclusions}

In this work we investigate the similarities and differences between resonant leptogenesis and leptogenesis through neutrino oscillations in the minimal extension of the standard model by two HNLs.
We find that the two mechanisms are closely related, and that the equations needed to describe the two mechanisms are in fact the same.
Since the defining feature of resonant leptogenesis, namely the resonant production of the baryon asymmetry is also present in leptogenesis via neutrino oscillations, we focus on the major difference between the two mechanisms, namely the question whether the majority of the BAU is produced during the \textit{freeze-in}, or \textit{freeze-out} of the heavy neutrinos.

We found significant overlap between the two regimes, namely, freeze-in leptogenesis turns out to play a major role in generating the BAU even for TeV and heavier Majorana neutrinos.
This regime mainly coincides with relatively large $\Delta M_N / M_N \sim 10^{-3}$ mass splitting,
compared to the one optimal for a resonant enhancement $\Delta M_N / M_N \sim 10^{-11}$.
Furthermore, the fact that the freeze-in regime extends large masses implies a strong dependence on the initial condition which was typically absent in resonant leptogenesis.

On the other hand, we also find that freeze-out leptogenesis remains viable for masses as low as $M = 5$~GeV.
This can be understood through the large decay efficiency of the HNLs, as a suppression factor of $M^2/T^2 \sim 10^{-3}$ is not sufficiently small to prevent baryogenesis.

Together, these two parametric regimes span all experimentally allowed masses for the heavy neutrinos, from a fraction of GeV, to $M_W$, and beyond.

\begin{acknowledgments}
\paragraph{Acknowledgments.} We thank Marco Drewes, Shintaro Eijima, Bj{\"o}rn Garbrecht, Jacopo Ghiglieri, Mikko Laine, and Apostolos Pilaftsis for helpful comments and discussions. This work was supported by ERC-AdG-2015 grant 694896 and by the Swiss National Science Foundation Excellence grant 200020B\underline{ }182864.
\end{acknowledgments}

\bibliography{lepto_refs}

\end{document}